\documentclass[sn-nature]{sn-jnl}

\RequirePackage{amsthm}

\usepackage{graphicx}%
\usepackage{multirow}%
\usepackage{amsmath,amssymb,amsfonts}%
\usepackage{amsthm}%
\usepackage{mathrsfs}%
\usepackage[title]{appendix}%
\usepackage{xcolor}%
\usepackage{textcomp}%
\usepackage{manyfoot}%
\usepackage{booktabs}%
\usepackage{algorithm}%
\usepackage{algorithmicx}%
\usepackage{algpseudocode}%
\usepackage{listings}%
\usepackage{url} 

\theoremstyle{thmstyleone}%
\theoremstyle{thmstyletwo}%

\theoremstyle{thmstylethree}%

\UseRawInputEncoding
\begin{document}

\title[Article Title]{Reentrant Phase Transition in Quasiperiodic Photonic Waveguides}


\author[1,4,5]{\fnm{Yang} \sur{Chen}}
\equalcont{These authors contributed equally to this work.}

\author[2,3]{\fnm{Ze-Zheng} \sur{Li}}
\equalcont{These authors contributed equally to this work.}

\author[1,4,5]{\fnm{Hua-Yu} \sur{Bai}}
\equalcont{These authors contributed equally to this work.}

\author[3]{\fnm{Shuai-Peng} \sur{Guo}}
\author[1,4,5]{\fnm{Tian-Yang} \sur{Zhang}}
\author[3]{\fnm{Xu-Lin} \sur{Zhang}}
\author[3]{\fnm{Qi-Dai} \sur{Chen}}
\author[1,4,5]{\fnm{Guang-Can} \sur{Guo}}
\author[1,4,5]{\fnm{Fang-Wen} \sur{Sun}}
\author*[3]{\fnm{Zhen-Nan} \sur{Tian}}\email{zhennan\_tian@jlu.edu.cn}
\author*[1,4,5]{\fnm{Ming} \sur{Gong}}\email{gongm@ustc.edu.cn}
\author*[1,4,5]{\fnm{Xi-Feng} \sur{Ren}}\email{renxf@ustc.edu.cn}
\author*[2,3]{\fnm{Hong-Bo} \sur{Sun}}\email{hbsun@tsinghua.edu.cn}

\affil[1]{\orgdiv{CAS Key Laboratory of Quantum Information}, \orgname{University of Science and Technology of China}, \orgaddress{\city{Hefei, Anhui}, \postcode{230026}, \state{China}, \country{Country}}}

\affil[2]{\orgdiv{State Key Laboratory of Precision Measurement Technology and Instruments, Department of Precision Instrument}, \orgname{Tsinghua University}, \orgaddress{\city{Beijing}, \postcode{100084}, \country{China}}}

\affil[3]{\orgdiv{State Key Laboratory of Integrated Optoelectronics, College of Electronic Science and Engineering}, \orgname{Jilin University}, \orgaddress{\city{Changchun, Jilin}, \postcode{130012},  \country{China}}}

\affil[4]{\orgdiv{CAS Center for Excellence in Quantum Information and Quantum Physics}, \orgname{University of Science and Technology of China}, \orgaddress{\city{Hefei, Anhui}, \postcode{230026}, \country{China}}}

\affil[5]{\orgdiv{Hefei National Laboratory}, \orgname{University of Science and Technology of China}, \orgaddress{\city{Hefei, Anhui}, \postcode{230088}, \country{China}}}


\abstract{Anderson transition in quasiperiodic potentials and the associated mobility edges have been a central focus in quantum simulation across multidisciplinary physical platforms. While these transitions have been experimentally observed in ultracold atoms, acoustic systems, optical waveguides, and superconducting junctions, their interplay between quasiperiodic potential and long-range hopping remains unexplored experimentally. In this work, we report the observation of localization-delocalization transition induced by the hopping between the next-nearest neighboring sites using quasiperiodic photonic waveguides. Our findings demonstrate that increasing the next-nearest hopping strength induces a reentrant phase transition, where the system transitions from an initially extended phase into a localized phase before eventually returning to an extended phase. This remarkable interplay between hopping and quasiperiodic potential in the lattice models provides crucial insights into the mechanism of Anderson transition. Furthermore, our numerical simulation reveals that this phase transition exhibits a critical exponent of $\nu \simeq 1/3$, which is experimentally observable for system sizes $L\sim10^3$ - $10^4$. These results establish a framework for direct observation of the Anderson transition and precise determination of its critical exponents, which can significantly advance our understanding of localization physics in quasiperiodic systems. }

\keywords{Reentrant phase transition, quasiperiodic systems, Anderson localization, femtosecond laser direct writing, critical phenomena}



\maketitle

\section{Introduction}\label{sec1}

Since its theoretical prediction by Philip W. Anderson in 1958, Anderson localization has remained one of the most fascinating concepts in condensed matter physics \cite{Evers2008Anderson, Lagendijk2009Fifty, dikopoltsev2022observation}. While it has been extensively studied in electronic systems and well-established in theory, its experimental observation has long been challenging due to the difficulties of direct imaging their wavefunctions. Furthermore, the critical exponents associated with this phenomenon remain poorly understood. In the past two decades, the development of advanced optical simulation platforms \cite{segev2013anderson, yu2021engineered} has renewed the interest in Anderson localization, offering a promising pathway to address this enduring challenge in physics.

In one-dimensional systems with random potentials, all states are known to exhibit localization. However, quasiperiodic potentials enable phase transitions from extended to localized states as the potential strength increases, exhibiting critical behavior at some critical potential strength. The most celebrated example is the Aubry-Andr{\'{\rm e}} (AA) model, originally derived from a two-dimensional lattice in a strong magnetic field \cite{Harper1955}, which shows a sharp transition at a critical strength $V_c$. More generally, quasiperiodic systems can display energy-dependent mobility edges (MEs), separating the extended states from the localized states in the spectrum. These MEs have been thoroughly investigated through analytical and numerical methods \cite{Biddle2010Pre, Ganeshan2015, Lin2024Fate}. Meanwhile, the Anderson transitions have been observed in ultracold atoms \cite{jendrzejewski2012three, semeghini2015measurement, An2018}, acoustic crystals \cite{Ni2019, Apigo2019}, photonic systems \cite{Lahini2009, Weidemann2022} , and topological insulators \cite{Stuetzer2018, Meier2018}, with promising extensions to many-body localization \cite{Schreiber2015, rispoli2019quantum, abanin2019many}. As a fundamental concept, this localization has enabled important applications in imaging \cite{karbasi2014image} and lasing \cite{Wiersma2008, Vynck2023Light}. 

From a theoretical perspective, the Hamiltonian comprises nearest-neighboring hopping and an on-site quasiperiodic potential, whose competition determines Anderson transitions and ME formation. Localized states occur when the on-site potential dominates, while extended states appear when the hopping term dominates. However, next-nearest-neighbor (NNN) hopping can dramatically alters this conventional wisdom in terms of reentrant transition. In this work, we experimentally demonstrate the localization-delocalization transition (LDT) and the reentrant transition in 1D quasiperiodic chains induced by the NNN hopping, using evanescently coupled waveguides. The on-site potential follows an interpolated Aubry-Andr{\'{\rm e}}-Fibonacci form \cite{Kraus2012a, Goblot2020}, and we observe that the NNN hopping drives a reentrant transition from an extended phase to a localized phase and finally to an extended phase. We characterize this phase transition using the inverse participation ratio (IPR) and show that the associated critical exponent $\nu \simeq 1/3$, which can be determined in future experiments with waveguide sizes $L \sim 10^3 - 10^4$. Our results demonstrate that the optical waveguides provide an ideal platform for investigating Anderson transitions 
and determining their critical exponents. This approach provides a foundation for a deeper understanding of localization phenomena and paves the way for future studies of complex quasicrystals \cite{wang2020localization}.

\section{Results}\label{sec2}

\subsection{Theoretical model and phase diagram}

\begin{figure}[t]%
\centering
\includegraphics[width=1.0\textwidth]{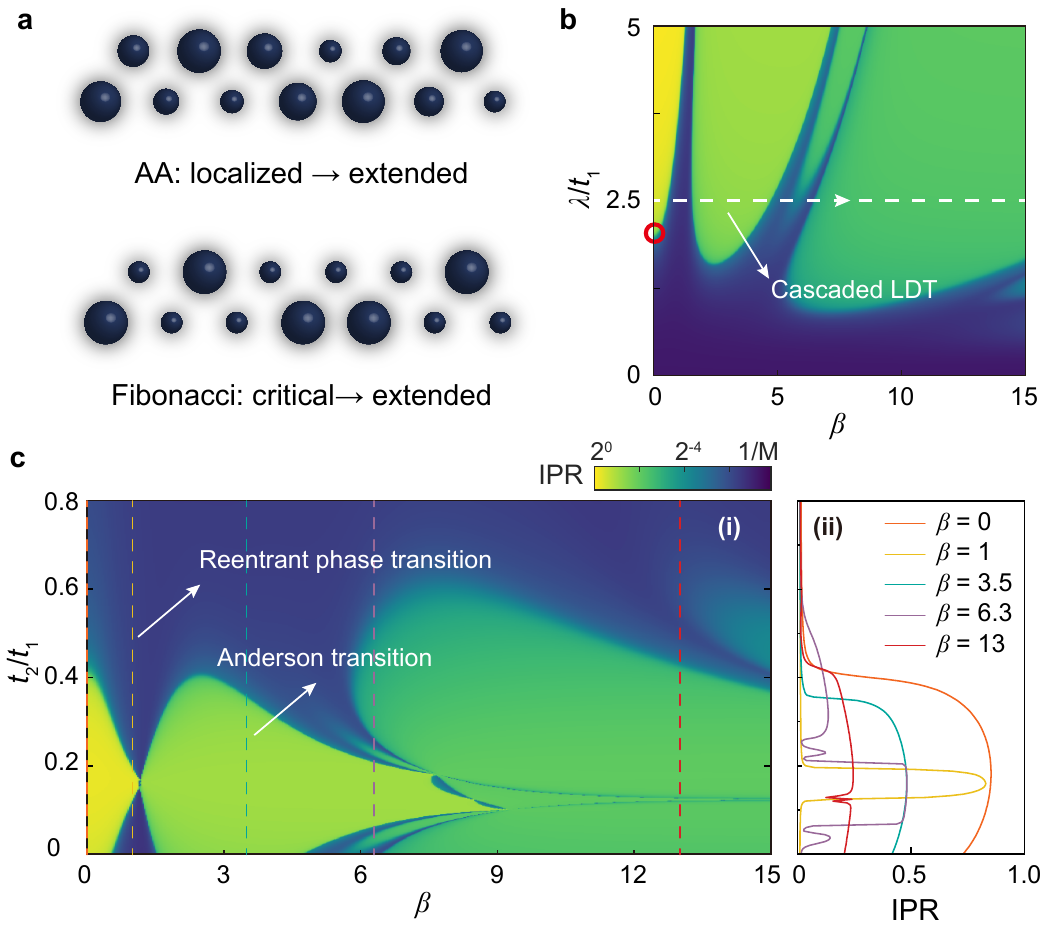}
\caption{\textbf{Localization-delocalization transition (LDT) induced by quasiperiodicity ($\beta$) and NNN hopping ($t_2$).} \textbf{a}, Schematics of two types of one-dimensional quasiperiodic zigzag chains: Aubry-Andr{\'{\rm e}} (AA, top) and Fibonacci (bottom). Here, the quasiperiodic potential $V_j$ is realized by the on-site energy, and the NNN hopping $t_2$ can be adjusted by varying the NNN lattice spacing. The IAAF model interpolates between the AA and Fibonacci modulation. \textbf{b}, Inverse participation ratio (IPR) of the ground state in the IAAF model. The white dashed line marks the LDT induced by $\beta$, and the red circle indicates the transition point of the AA model with $\lambda/t_1=2$. \textbf{c}(i), IPR of the IAAF model at $\lambda/t_1=2.5$. The yellow ($\beta=1$) and purple ($\beta=6.3$) lines show the reentrant transition induced by NNN hopping, while the remaining lines show the Anderson transition. \textbf{c}(ii), Ground state IPR as a function of $t_2$ for the five vertical lines shown in \textbf{c}(i).}\label{fig1}
\end{figure}

Two canonical quasiperiodic models, the AA and Fibonacci models, are topologically equivalent and nontrivial \cite{Kraus2012a}. Recent studies have revealed that the interplay between quasiperiodicity \cite{Roy2021}, interactions \cite{An2021, gonccalves2024incommensurability}, and non-hermiticity \cite{Longhi2019, Weidemann2022, Lin2022}, as well as the inclusion of short- and long-range hoppings \cite{Biddle2010Pre, Vaidya2023}, can lead to the coexistence of localized and extended phases in lower-dimensional systems. Given these findings, we investigate the Hamiltonian of a generalized model with an IAAF potential, expressed as 
\begin{equation}
    H = \sum_{i=1}^{L}\lambda V_i(\beta) c_i^\dagger c_i + \sum_{i,j}(t_{ij} c_{i}^\dagger c_j + \text{h.c.})
    \label{eq5}
\end{equation}
where $c_i^\dagger (c_i)$ is the creation (annihilation) operator at site $i$, $L$ is the chain length, and $h.c.$ is the Hermitian conjugate. The hopping amplitudes are defined as $t_{ij} = t_{|i-j|}$ for $i-j = \pm 1, \pm 2$, and $t_{ij} = 0$ otherwise, with $t_2$ characterizing the NNN hopping. $\lambda$ is the tuning quasiperiodic strength, and the on-site quasiperiodic potential modulation is given by
\begin{equation}
    V_i(\beta) = -\frac{\tanh[\beta (\cos(2\pi bi+\phi)- \cos \pi b)]}{\tanh \beta}
    \label{eq2}
\end{equation}
where $b$ is irrational, describing a quasiperiodic system, and $\phi$ is a phase shift. The quasiperiodicity is controlled by $\beta$, interpolating between the AA model ($\beta\rightarrow0$) and the Fibonacci model ( $\beta\rightarrow\infty$).

We choose $b = (\sqrt{5}-1)/2$ and focus on the regime $\lambda/t_1 > 2$, where the on-site term dominates, yielding a tendency to localized states. The phase $\phi$ is set to 0, as it does not affect the core physics under consideration. However, $\phi$ remains significant in other contexts, such as engineering edge-state pumping in quasicrystals \cite{citro2023thouless}. To characterize the LDT in the IAAF model, we calculate the inverse participation ratio (IPR) for each normalized state $\left|\psi_n\right> =(\phi_n^1, \phi_n^2, \cdots, \phi_n^L)^T$ as ${\rm IPR}_n = \sum_{i=1}^{L} |\phi_n^i|^4$. For an extended state, $\text{IPR}\sim \mathcal{O}(1/L
)$; whereas for a localized state, $\text{IPR}\sim \mathcal{O}(1)$. This ratio depends solely on the magnitude of wavefunctions at each site and can also be applied to the time evolution dynamics \cite{Wang2022Edge}. 

\begin{figure}[t]%
\centering
\includegraphics[width=1.0\textwidth]{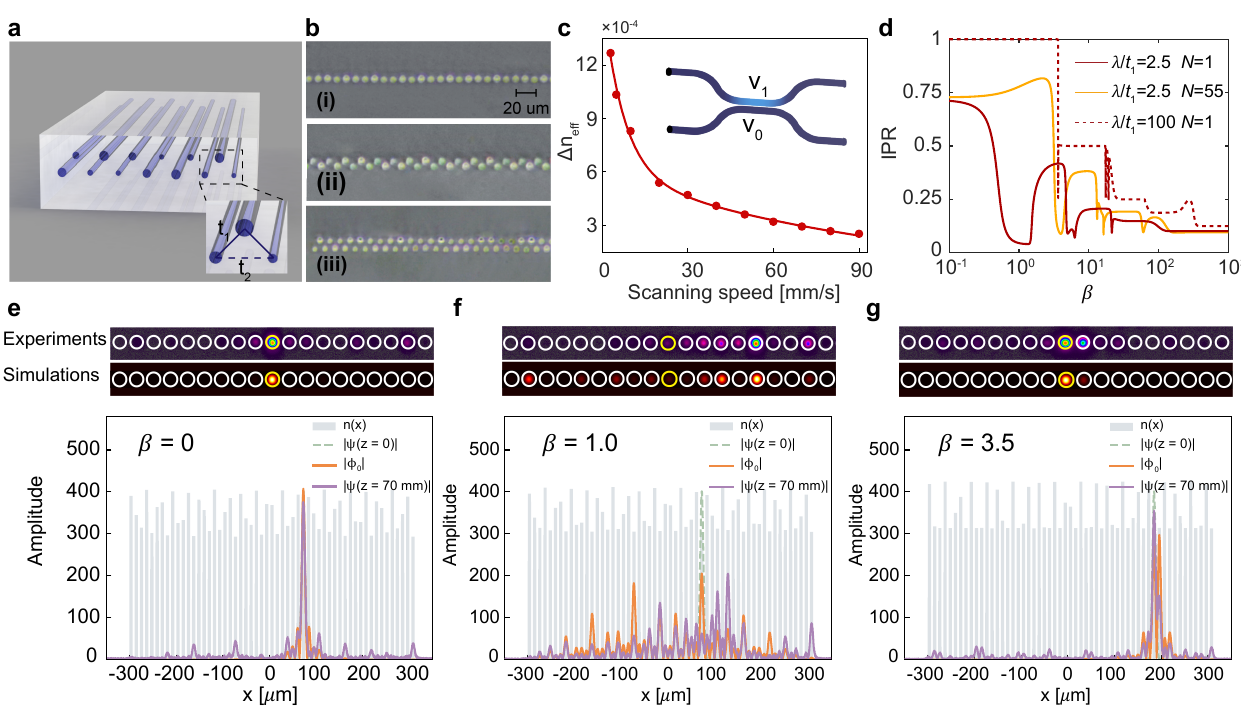}
\caption{\textbf{Experimental setup and LDT measurements induced by quasiperiodicity.} \textbf{a}, Schematic of the one-dimensional zigzag-type quasiperiodic waveguide array, where the effective refractive index of each waveguide is controlled by adjusting the laser's scanning speed. \textbf{b}, End-facet microscopy image of the IAAF waveguide array. The nearest-neighbor coupling $t_1$ remains fixed, while the NNN coupling $t_2$ is tuned via the NNN waveguide spacing. \textbf{c}, Effective refractive index change ($\Delta n_{\text{eff}}$, relative to the surrounding medium) versus fabrication scanning speed. \textbf{d}, IPR for the 55-site IAAF chain, showing the $N=1$ lowest-energy eigenmodes (red: $\lambda/t_1=2.5$ solid, 100 dashed) and the $N=55$ highest-energy eigenmodes ($\lambda/t_1=2.5$, yellow solid) versus $\beta$. \textbf{e}-\textbf{g}, When light is injected into the waveguide (yellow circle) and $\beta$ is varied (with $t_2\sim 0$), the measured and simulated output patterns at $z=70$ mm demonstrate three distinct regimes: \textbf{e}, localization at $\beta=0$ (input position: $34$-th waveguide), \textbf{f}, delocalization at $\beta=1.0$ ($34$-th), \textbf{g}, re-localization at $\beta=3.5$ ($44$-th). The bottom row in \textbf{e}-\textbf{g} shows the corresponding effective refractive index distribution, simulated eigenmodes, and normalized wavefunction amplitudes at $z=0$ mm and $z=70$ mm.} \label{fig2}
\end{figure}

We begin by analyzing the IPR of the ground state (GS) in the IAAF model (with $t_2 = 0$) as a function of $\beta$ and $\lambda/t_1$. The phase diagram is presented in Fig. \ref{fig1}b. For $\beta \rightarrow 0$, $V_i \rightarrow \cos(\pi b)- \cos(2\pi b i + \phi)$, $H$ is reduced to the AA model, and the phase transition from extended to localized states occurs at $\lambda/t_1 = 2$, as predicted by self-duality \cite{aubry1980analyticity}. In the $\beta\rightarrow\infty$ limit, the potential transitions to the Fibonacci model, yielding critical states with self-similar wavefunctions \cite{kohmoto1987critical}. Increasing $\beta$ induces a series of cascade transitions from extended ($\lambda/t_1 < 2$) or localized states ($\lambda/t_1 > 2$) to critical states. For $\lambda/t_1 > 2$, the multiple phase transition is characterized by IPR plateaus. In each plateau, as $t_1\rightarrow0$, the IPR is half that of the previous plateau (Fig. \ref{fig2}d), corresponding to a state localized over 1, 2, 4, $\cdots$ sites. This cascade behavior was recently observed in experiments using cavity-polariton devices for $L >50 $ \cite{Goblot2020}. 

Introducing NNN hopping can significantly modify the phase diagram. We present the phase diagram in Fig. \ref{fig1}c, depicting the IPR as a function of $\beta$ and $t_2/t_1$ for fixed $\lambda = 2.5t_1$. For $t_2/t_1<0.16$, the cascade transitions with increasing $\beta$ persist. However, a novel reentrant phenomenon emerges: for fixed $\beta$, increasing $t_2$ leads to extended-localized-extended transitions, driven by the complex interplay between quasiperiodicity and NNN hopping. This reentrant behavior is clearly illustrated in the calculated IPR, as shown in Fig. \ref{fig1}c(ii). Notably, reentrant transitions have been experimentally observed in time-varying quasiperiodic systems \cite{shimasaki2024anomalous} and theoretically predicted in static quasiperiodic systems \cite{Roy2021, Vaidya2023}. From a much broader perspective, the NNN hopping, even with a magnitude much smaller than nearest-neighbor hopping, can significantly influence its physics, leading to nontrivial phenomena \cite{Defenu2023Long}. For instance, recent studies demonstrate that the NNN hopping is crucial for realizing topological phases like the Haldane phase \cite{jotzu2014experimental, zhao2024realization}, and for addressing the long-standing puzzle of high-temperature superconductivity in many-body systems \cite{jiang2019superconductivity, Hao2024coexistence}.

\noindent\textbf{Experimental results} 

Our experimental system consists of evanescently coupled optical waveguides arranged in a $55$-site zigzag chain (Fig. \ref{fig2}a), fabricated using the Femtosecond laser direct writing technique. The diffraction of light in this paraxial waveguide array is described by the Schr\"odinger-type equation
\begin{equation}
    i\partial_z\psi\left(x, y, z\right) = -{\frac{1}{2k_0}} \nabla_{\perp}^2\psi\left(x, y, z\right) -{\frac{k_0 \Delta n(x,y)}{n_0}}\psi\left(x, y, z\right)
\end{equation}
where $\psi\left(x, y, z\right)$ is the scalar optical mode electric field amplitude defined by $E\left( x, y,z\right) = \psi\left(x, y, z\right) \exp\left(ik_0z-i\omega t \right)$, $E$ is the electric field, $k_0=2\pi n_0/\lambda$ is the wave number, $n_0$ the background refractive index, $\omega = 2\pi c/\lambda$ the optical frequency, $\lambda$ the wavelength in vacuum, $c$ the speed of light, $\nabla_\perp^2$ the transverse Laplacian in the $(x,y)$ plane and $\Delta n(x,y)$ the local refractive index change in the waveguide relative to the background medium. For single-mode waveguides, the propagation dynamics can be effectively described by the tight-binding model \cite{longhi2009quantum, chen2021tight}, where the hopping amplitude and on-site potential are controlled by the inter-waveguide spacing and the effective refractive index modulation, respectively. 

In our design, the nearest-neighbor waveguide distance $d$ is fixed, and the NNN hopping is achieved by adjusting the relative distance between the NNN waveguides (Fig. \ref{fig2}a-b). The effective refractive index of each waveguide is adjusted according to Eq. \ref{eq2} by controlling the scanning speed of the writing laser, as shown in Fig. \ref{fig2}c (see methods).  For the 1D waveguide arrays, the cascade transition predicted by the model are shown in Fig. \ref{fig2}d (IPR plateaus, theory) and Fig. \ref{fig2} e-g (experiments, $d=11.2~\mu$m). By increasing $\beta$, the system transitions from a one-site localized state to an extended state, and subsequently to a two-site localized state. In the experiment, a coherent laser is injected into the waveguide where the GS is localized, and the light field pattern is observed at a propagation distance of $z = 70$ mm.

\begin{figure}[t]%
\centering
\includegraphics[width=1.0\textwidth]{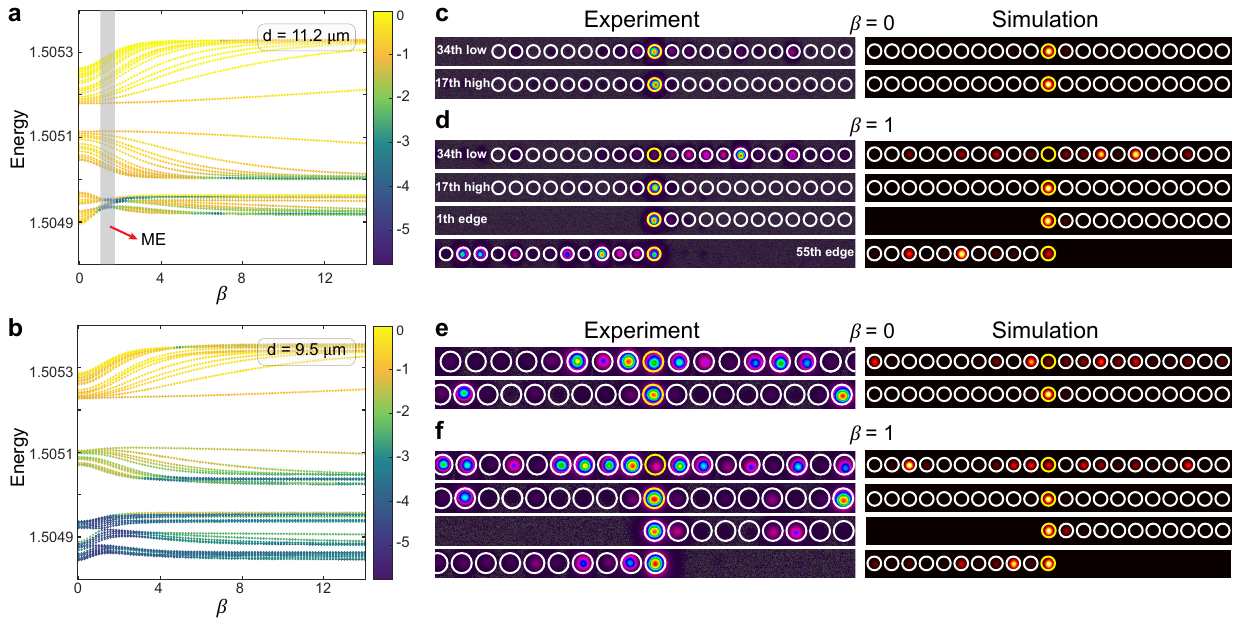}
\caption{\textbf{Experimental observation of LDT and mobility edge (ME) in IAAF waveguide arrays.} \textbf{a}, IPR of all eigenmodes for nearest-neighbor waveguide distance $d=11.2~\mu$m, with gray shading indicating ME emergence. \textbf{b}, IPR of all eigenmodes for $d=9.5~\mu$m. \textbf{c}, Measured and simulated output intensity at $z = 70$ mm for $\beta=0$, $d=11.2~\mu$m. The observed light localization at all input positions (yellow circles) agrees with the AA model prediction. \textbf{d}-\textbf{f}, Results for parameter sets $\{\beta,~d\}=\{1.0,~11.2~\mu\text{m}\}, ~\{0,~9.5~\mu\text{m}\}$, and $\{1.0,~9.5~\mu\text{m}\}$. The eigenenergy is characterized by the propagation constant (in units of $1/k_0$).}\label{fig3}
\end{figure}

For the above continuous model, its tight-binding description \cite{chen2021tight} is not straightforward. Thus, we first experimentally validate the tight-binding approximation, with results for $d= 9.5$ and 11.2 $\mu$m presented in Fig. \ref{fig3}. The IPR of eigenmodes in the optical waveguide arrays is calculated using the finite-element method (Supplementary Section II). For $d = 11.2$ $\mu$m ($\lambda/t_1 = 3.9$, $t_1 = 0.35~ \text{mm}^{-1}$, Fig. \ref{fig3}a), all states are localized at $\beta =0$ (AA limit). As $\beta$ increases, the GS and low-lying states transition to extended states, while the high-energy states remain localized, resulting in mobility edges (MEs). In contrast, for $d = 9.5$ $\mu$m ($\lambda/t_1 = 2.5$, $t_1 = 0.55~ \text{mm}^{-1}$, Fig. \ref{fig3}b), the system shows distinct behavior from tight-binding predictions (which is the same as Fig. \ref{fig3}a) due to the influence of $t_2$ hopping, $\lambda/t_1$, and continuous spatial distribution of the waveguide eigenmodes. Here, while the states are localized at $\beta = 0$ (tight-binding prediction), the low-lying states exhibit extended behavior in a finite chain ($L = 55$) when the localization length $\xi > L$ (see Supplementary Section II for details). These results are confirmed by wave packets evolution (Fig. \ref{fig3}c-f for both $d$ at $\beta =0$ and $1$), showing excellent agreement between experiment and simulation. The measurements show that localization in the waveguide array is significantly influenced by both $t_2$ and $\lambda/t_1$ \cite{Biddle2011}, even at $t_2/t_1 \ll 1$ (0.05 at $d = 9.5$ $\mu$m; 0.03 at $d=11.2$ $\mu$m).
\begin{figure}[t]%
\centering
\includegraphics[width=1.0\textwidth]{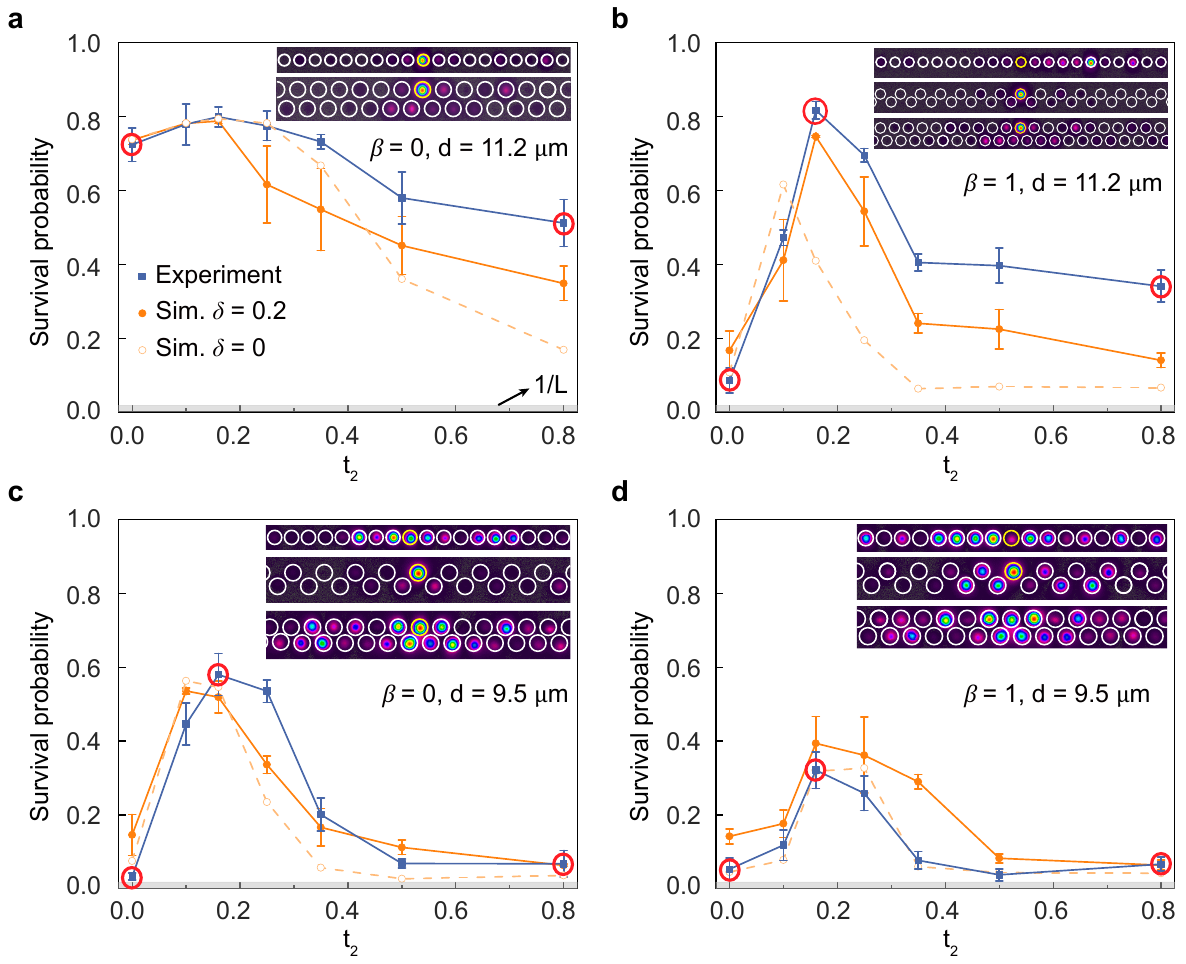}
\caption{\textbf{Delocalization and reentrant behavior induced by NNN hopping.} \textbf{a}, Survival probabilities (SPs) for $\beta=0$, and $d=11.2~\mu$m: experimental measurements (blue dots, averaged over five propagation lengths from $30-70$ mm in 10 mm increments), numerical simulations (orange dots for results averaged over five sets of disorder with strength $\delta=0.2$; light orange dots for $\delta=0$). \textbf{b}-\textbf{d}, Corresponding SPs for \textbf{b}, $\beta=1.0$, $d=11.2~\mu$m; \textbf{c}, $\beta=0$, $d=9.5~\mu$m; \textbf{d}, $\beta=1.0$, $d=9.5~\mu$m. Insets show measured output light intensity at $z = 70$ mm. Error bars indicate standard error of five lengths or disorder realizations ($\overline{\text{SP}} = (\sum_{i}\text{SP}_i)/5$).}\label{fig4}
\end{figure}

Furthermore, Fig. \ref{fig1}c shows that increasing $t_2$ coupling induces reentrant phenomena (e.g., $\beta=1.0$). Such effects are unrealizable in conventional one-dimensional waveguide arrays due to the exponential decay of evanescent coupling between distant waveguides. To overcome this difficult, we design zigzag-structured waveguide arrays with $d = 9.5$ or 11.2 $\mu$m. Two-port directional couplers are fabricated to determine $t_1$ and $t_2$ (see Methods). In principle, this configuration enables $t_2 > t_1$, allowing us to access to regime with much richer physical behaviors. By fabricating different zigzag waveguide arrays, we explore the phase diagram by tuning $\beta$ and $t_2/t_1$ at a fixed $\lambda/t_1$. 

The reentrant phase transition is one of the most important features in this model. Fig. \ref{fig4} presents the output intensity averaged over five propagation lengths. The coupling period is $T = \pi/t_1$, and the maximum periods are $12.2T$ ($d=9.5$ $\mu$m) and $7.8T$ ($d=11.2$ $\mu$m). These lengths are sufficient to observe the reentrant behavior, with localization properties characterized by the averaged survival probability (SP). For the localized states, the SP remains a constant over different propagation lengths, whereas for the extended states, the SP will finally approach $1/L$ \cite{Wang2022Edge}. This behavior is experimentally demonstrated in Fig. \ref{fig4}. For the AA model with $\beta = 0$, and $d = 11.2$ $\mu$m (Fig. \ref{fig4}a), all states are localized, with SP values ranging from $0.5$ to $0.8$. As $t_2$ increases, the wave packets remain localized even at $t_2/t_1 = 0.8$. In contrast, for $\beta = 1$ (Fig. \ref{fig4}b), the wavefunctions initially exhibit extended behavior at $t_2/t_1 \sim 0$, become localized near $t_2/t_1 \sim 0.2$, and finally re-enter an extended phase for $t_2/t_1 > 0.4$. The results for $d = 9.5$ $\mu$m, shown in Fig. \ref{fig4}c ($\beta = 0$) and Fig. \ref{fig4}d ($\beta = 1$), reveal a similar reentrant transition. In all cases, the SP peak observed at $t_2/t_1 \simeq 0.16$ agrees well with the phase diagram shown in Fig. \ref{fig1}c(ii). However, it should be emphasized that the numerical simulations of the continuous waveguide arrays in Fig. \ref{fig4} show slight discrepancies with the experimental results, which arise from the fabrication uncertainties in the waveguides. To address this issue, we introduce random refractive index fluctuations in each waveguide, constrained within a finite range characterized by the disorder strength $\delta$ (defined relative to the quasiperiodic modulation’s refractive index variation, see Supplementary Section VI). The results are shown as orange dots in the same figure. The observed slight deviations between experimental and numerical results are well explained by introducing a weak random potential.

\begin{figure}[t]%
\centering
\includegraphics[width=1.0\textwidth]{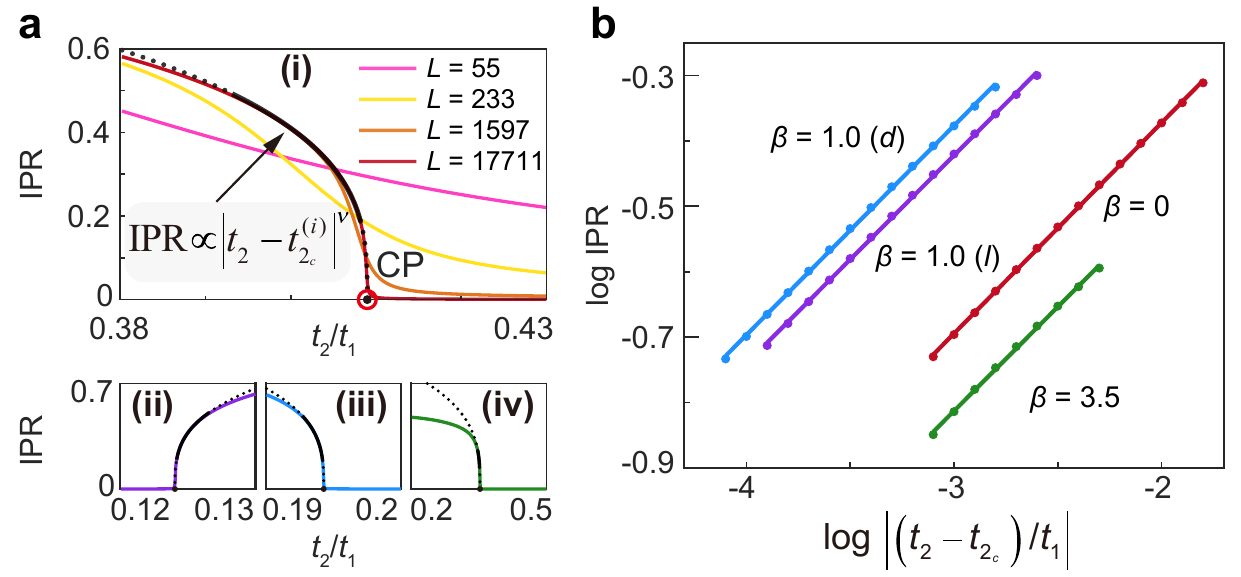}
\caption{\textbf{Critical properties of LDT in the IAAF model.} \textbf{a}, IPR of the lowest-energy eigenmode versus $t_2$ for: (i), system sizes $L=55,~233,~1597,~17711$ at $\beta = 0$. (ii)-(iv), $L=17711$ at $\beta=1.0$ (purple,~localization transition), $\beta=1.0$ (light blue, delocalization transition), and $\beta=3.5$ (green). Black lines correspond to the power-law fits IPR$\sim (t_2-t_{2_c})^\nu$ near critical points $t_{2_c}$, where the solid black lines indicate the fitted area. \textbf{b}, Log-log plot of IPR as a function of $t_2-t_{2_c}$ for different LDT processes.}\label{fig5}
\end{figure}

Finally, we address the observed LDT and its connection with the critical behavior. The LDT is particularly significant as it represents a phase transition rather than a trivial (smooth) crossover process. Our experiments demonstrate a nontrivial reentrant transition between extended and localized states as $t_2/t_1$ increases, establishing NNN coupling as a fundamental control parameter for localization properties. To determine whether this transition constitutes a phase transition, we analyze the IPR of the GS at different system sizes ($L=55, ~233, ~1597, \text{and} ~17711$). In the thermodynamic limit, the IPR serves as an order parameter exhibiting universal scaling behavior at different $\beta$. For fixed $t_1$, we determine the critical behavior through the scaling relation \cite{Evers2008Anderson, Yao2019}: 
\begin{equation}
\text{IPR} \propto |t_2 -t_{2_c}|^\nu,
\end{equation}
where $\nu$ is the critical exponent. For the cases shown in Fig. \ref{fig5}a(i)-(iv), the critical point $t_{2_c}$ are $0.409$, $0.124$, $0.194$ and $0.353$, respectively. From Fig. \ref{fig5}b, we extract the critical exponent $\nu$ as $0.322\pm0.003$, $0.318\pm0.004$, $0.320\pm0.004$, and $0.318\pm0.010$. This exponent is expected to be extracted experimentally in future studies. 

\noindent\textbf{Discussion and Conclusion} 

Let us briefly discuss the future of simulation in physics. While quantum and classical simulations have been widely employed to study phase transitions in various systems \cite{Georgescu2014quantum, daley2022practical}, finite-size effects have thus far greatly limited the precise determination of the critical exponents. Our numerical simulations (Fig. \ref{fig5}) indicate that systems with $L \sim 10^4$ sites are sufficient to quantitatively determine the critical phenomena and extract the corresponding exponents. Although current experiments typically use systems of size $L \sim 10 - 100$, recent advances in 3D optical waveguide arrays ($49 \times 49$ sites \cite{tang2018experimental}) and 2D optical lattices (15,000 microtraps \cite{Gerritsma2007lattice}) suggest the feasibility of scaling to larger systems. We anticipate that future advances in fabrication and measurement techniques will enable the study of much larger systems, yielding the precision required to determine the phase transitions and critical exponents. 

To conclude, we experimentally demonstrate the LDT in quasiperiodic optical systems induced by the interplay between quasiperiodicity and NNN hopping. In a much larger system, we show that this transition exhibits a scaling law near the critical point, with a critical exponent $\nu \simeq 1/3$, which is the unique signature of a phase transition. These findings offer key insights into critical phenomena in disordered and quasiperiodic systems. Our results pave the way for further exploration of LDT and criticality in higher-dimensional quasiperiodic systems with long-range hopping \cite{wang2020localization}, and potential novel physics emerging from nonlinear interactions \cite{Alvarez2015, Xia2021nonlinear, Marius2023quantized}, which remain open questions. Moreover, as system sizes continue to increase, this simulation platform may emerge as a powerful experimental tool for discovering new physical phenomena.

\section{Methods}\label{sec11}

\textbf{Experimental setup}

\noindent The quasiperiodic waveguide arrays are fabricated using the femtosecond laser direct writing technique. A 1030 nm laser (with a repetition rate of $1$ MHz and pulse duration of $290$ fs) is focused inside the borosilicate glass (Corning Eagle XG). We employ a previously developed multi-foci-shaped femtosecond pulsed (MFSFP) method for all waveguide structures, utilizing a spatial light modulator (SLM, X13138, Hamamatsu Photonics) to modulate the femtosecond pulses and an objective lens ($40\times$, N.A.=$0.75$, Olympus) for focusing. The glass chip is mounted on a high-precision three-axis Aerotech motion stage (ABL1000), which finely tunes the translation speed and the waveguide trajectory. To capture long-time evolution dynamics, the total glass length is set at 75 mm. A diamond wire saw (STX-202A, Shenyang Kejing) is used to cut the chip, followed by polishing of the end facets. The mode fields and the energy distribution of the end facets of the arrays are analyzed using a beam analyzer (SP928, Spiricon Inc.). 

In our experiment, we fabricate single-mode waveguides operating at 808 nm. First, a reference waveguide is created with a scanning speed of $50$ mm/s and a laser power of $410$ mW, resulting in a refractive index increase of $3.61\times10 ^{-4}$ relative to the surrounding medium. The transverse and longitudinal dimensions of the waveguide are measured at $4.1$ $\mu$m and $4.0$ $\mu$m, respectively. To achieve refractive index modulation, we vary the scanning speed of the translation stage while maintaining a constant laser power of $410$ mW, which was determined through preliminary experiments. The intensity of the laser power affects the refractive index range and could lead to undesired multimode characteristics in the waveguide. The relative effective refractive index change ranges from 2.52$\times10^{-4}$ to 1.27$\times10^{-3}$, corresponding to laser scanning speeds from 90 mm/s to 3 mm/s (Fig. \ref{fig2}c), enabling the realization of on-site quasiperiodicity modulation.  

\noindent\textbf{Waveguide optical properties characterization}

\noindent The effective refractive index of the waveguide is measured at various scanning speeds by constructing an asymmetric directional coupler (ADC) (see Supplementary Information). The transmission for this ADC is given by 
\begin{equation}
    {\rm Trans} = -{\frac{\kappa^2}{\kappa^2+\Delta^2}} \sin^2(\sqrt{\kappa^2+\Delta^2}\cdot{z})
\end{equation}
\begin{equation}
    \Delta = {\frac{\pi (n\mathrm{_{eff_0}}-n\mathrm{_{eff_1})}}{\lambda_{0}}}
\end{equation}
where $\kappa$, $z$, $n\mathrm{_{eff_0}}$, $n\mathrm{_{eff_1}}$ and $\lambda_{0}$ are the coupling coefficient, coupling length, effective refractive index of the reference waveguide, effective refractive index of the measured waveguide, and the wavelength of light in a vacuum, respectively. Thus, the effective refractive index of the measured waveguide, $n\mathrm{_{eff1}}$, can be obtained by characterizing the transmission of the ADC. The relation between the effective refractive index and scanning speed is presented in Fig. \ref{fig2}c. Due to the asymmetrical design of the waveguide, the coupling varies at different spatial angles. In the Supplementary Section II, we illustrate the waveguide coupling at various angles relative to the waveguide spacing, as measured using directional couplers of different coupling lengths.

\backmatter

\section*{Acknowledgments}

This research is supported by the National Key Research and Development Program of China (2022YFA1204704), the National Natural Science Foundation of China (Nos. T2325022, U23A2074, 12204462), the CAS Project for Young Scientists in Basic Research (No.253 YSBR-049), the Strategic Priority Research Program of the Chinese Academy of Sciences (Grant No. XDB0500000), Key Research and Development Program of Anhui Province (2022b1302007), the Innovation Program for Quantum Science and Technology (2021ZD0303200, 2021ZD0301200, 2021ZD0301500), and the Fundamental Research Funds for the Central Universities (WK2030000107, WK2030000108). This work was partially carried out at the USTC Center for Micro and Nanoscale Research and Fabrication.

\section*{Author contributions} 
Y.C., Z.-Z.L., and H.-Y.B. contributed equally to this work. Y.C. conceived the idea and the experiment. Z.-Z.L. and S.-P.G. fabricated the photonic waveguides and performed the experiments, Y.C. and H.-Y.B. conducted numerical simulations and analysed the data under the supervision of X.-F.R., M.G., Z.-N.T., and H.-B.S. Y.C., Z.-Z.L., H.-Y.B., and M.G. wrote the manuscript, with input from all the authors.

\section*{Additional information}
\begin{itemize}
\item \textbf{Supplementary information} The online version contains supplementary material available. Correspondence and requests for materials should be addressed to X.-F.R., M.G., Z.-N.T., and H.-B.S..
\item \textbf{Competing interests} The authors declare no competing interests.

\end{itemize}


\bibliography{quasi-bibliography}

\end{document}